\documentclass[aps,prb,reprint,superscriptaddress,longbibliography,floatfix]{revtex4-2}
\usepackage[T1]{fontenc}
\usepackage[utf8]{inputenc}
\usepackage{amsmath,amssymb,bm,graphicx,booktabs}
\usepackage[colorlinks=true,citecolor=blue,linkcolor=blue,urlcolor=blue]{hyperref}

\begin{document}
\raggedbottom
% \title{Boundary-Geometry-Selective Higher-Order Topology from \texorpdfstring{$d_{xy}$}{dxy}-Wave Altermagnetic Proximity in Topological Insulators}
\title{Geometry-Selective Higher-order topological insulators induced by $d_{xy}$ altermagnet}
% Placeholder explicitly requested by the user; replace for an actual submission.
\author{Siyuan Cheng}
\affiliation{Department of Physics, Jiangsu University, Zhenjiang, 212013, China}

\author{Yuee Xie}
\affiliation{Department of Physics, Jiangsu University, Zhenjiang, 212013, China}  

\author{Xiaosen Yang} \altaffiliation{yangxs@ujs.edu.cn} 
\affiliation{Department of Physics, Jiangsu University, Zhenjiang, 212013, China}  

\author{Yuanping Chen} \altaffiliation{chenyp@ujs.edu.cn}
\affiliation{Department of Physics, Jiangsu University, Zhenjiang, 212013, China}  

\date{\today}
\begin{abstract}
We show that $d_{xy}$-wave altermagnetic proximity can induce boundary-geometry-dependent higher-order topological phases. The boundary geometry plays a crucial role in determining the topological properties: axis-aligned edge states remain gapless, whereas diagonal edges acquire effective masses with alternating signs, giving rise to four corner modes localized at the mass domain walls.  By projecting onto the edge-state subspace, we derive the effective edge Hamiltonians for different boundary geometries. Furthermore, $\pi$-flux defects introduced under periodic boundary conditions yield zero-energy bound states localized at the flux cores, providing complementary evidence for nontrivial topology without physical boundaries. The combination of the altermagnetic order and the Zeeman field induces a topological transition through the closing and reopening of the edge gap, while the bulk gap remains open. These findings highlight the crucial role of boundary geometry in higher-order topology and establish proximity-induced $d_{xy}$-wave altermagnetism as a route to engineering higher-order topological phases.
\end{abstract}
\maketitle

\section{Introduction}
The recognition of altermagnetism as a distinct class of collinear magnetic order has broadened the conventional understanding of ferromagnetism and antiferromagnetism~\cite{Libor2020,PhysRevB.102.014422,PhysRevX.12.031042,PhysRevX.12.040501,Bai2024,songAltermagnetsNewClass2025d,duanAntiferroelectricAltermagnetsAntiferroelectricity2025a,zhuTwoDimensionalFerroelectricAltermagnets2025,zhangCrystalsymmetrypairedSpinValley2025,yanNearlyTopologicalFlat2026,maryamLightinducedTopologicalPhase2026,liEnhancement$d$wavePairing2026}. Unlike conventional compensated antiferromagnets, altermagnets exhibit momentum-dependent spin splitting~\cite{xiaoSpinSpaceGroups2024,mcclartyLandauTheoryAltermagnetism2024a,kannanCorrelationdrivenMetalinsulatorTransition2026} governed by crystal symmetries while maintaining zero net magnetization~\cite{jiangEnumerationSpinSpaceGroups2024a,chenEnumerationRepresentationTheory2024,panOrbitalAltermagnetismTwo2026,fuMultipleTopologicalPhases2026,liEvolutionMagneticCorrelation2026,yarmohammadiSpinPolarizationEngineering2026,gonzalezTopologicalChiralMagnons2026}. A range of candidate materials and related systems, including MnTe~\cite{aminNanoscaleImagingControl2024a,gonzalezbetancourtAnisotropicMagnetoresistanceAltermagnetic2024,leeBrokenKramersDegeneracy2024a}, RuO$_2$~\cite{zhuAnomalousAntiferromagnetismMetallic2019,baiObservationSpinSplitting2022a,fedchenkoObservationTimereversalSymmetry2024,gqhb-2h45}, KV$_2$Se$_2$O~\cite{jiangMetallicRoomtemperatureDwave2025,5bns-qkgl,159t-fq4k}, Mn$_5$Si$_3$~\cite{leiviskaAnisotropyAnomalousHall2024,baduraObservationAnomalousNernst2025,reichlovaObservationSpontaneousAnomalous2024a}, and CrSb~\cite{reimersDirectObservationAltermagnetic2024a,zhouManipulationAltermagneticOrder2025b,yangThreedimensionalMappingAltermagnetic2025,wangNeelvectordependentMagnetoopticalSpin2026}, have been investigated theoretically and experimentally. Depending on the underlying crystal and magnetic symmetries~\cite{sunAndreevReflectionAltermagnets2023b,brekkeMinimalModelsTransport2024a,antonenkoMirrorChernBands2025a}, altermagnets can exhibit various momentum-dependent spin-splitting patterns, such as $d$-wave~\cite{jiNonHermitianHigherorderTopological2026,zhaoTailoringCornerStates2026,7dmx-h9h3,g1wx-v888,xyjn-8dqk}, $g$-wave~\cite{huoAltermagnetismItsInduced2026,zpbm-qr7d,15ct-lzds}, and $i$-wave~\cite{PhysRevB.111.125420,bg6g-98m1,fnwg-lfzg} forms, which can give rise to diverse topological and transport phenomena~\cite{ouassouDcJosephsonEffect2023a,zhuTopologicalSuperconductivityTwodimensional2023,linCoulombDragAltermagnets2025b,cheraghchiFloquetengineeredChernInsulator2026,liuLightinducedOddparityAltermagnets2026}. These distinct spin-splitting patterns give rise to various physical phenomena, including the anomalous Hall effect~\cite{fengAnomalousHallEffect2022a,satoAltermagneticAnomalousHall2024a,chenProbing$mathbitk$SpaceAlternating2025,hanMagnetizationfreeQuantumAnomalous2026}, unconventional proximity effect~\cite{zhuAltermagneticProximityEffect2026,alamProximityinducedSuperconductivityEmerging2026}, and open a new avenue for designing quantum materials.

Higher-order topology extends the conventional classification of topological phases by supporting topological states at boundaries of reduced dimensionality~\cite{hasanColloquiumTopologicalInsulators2010a,chiuClassificationTopologicalQuantum2016b,bansilColloquiumTopologicalBand2016a,dkvp-41zt,yanHigherOrderTopologicalOddParity2019a,PhysRevResearch.2.043300,xieHigherorderBandTopology2021,wangRealizationThreedimensionalPhotonic2025,luoHigherOrderTopologicalCorner2019a,liFloatingEdgeBands2025a,yangNonHermitianSecondorderTopological2026}. Higher-order topological phases support states localized at lower-dimensional boundaries, such as one-dimensional hinge modes~\cite{benalcazarElectricMultipoleMoments2017a,yanMajoranaCornerHinge2019a,liMajoranaCornerModes2023a,zhaoTopologicalElectronicStructure2023} and zero-dimensional corner modes~\cite{royMixedparityOctupolarPairing2021,renEngineeringCornerStates2020a,rlk2-psxm}. The integration of altermagnetic order with topological insulators offers a powerful platform for engineering higher-order topological phases~\cite{ghorashiAltermagneticRoutesMajorana2024a,liRealizingTunableHigherorder2024,guoHigherOrderTopologicalStates2025}. A common mechanism for realizing higher-order topology is to gap the edge states of a parent topological phase using mass terms whose signs differ between adjacent edges~\cite{kudoHigherOrderTopologicalMott2019a,PhysRevLett.121.096803,PhysRevLett.122.076801,jiNonHermitianSecondorderTopological2024,wanHigherorderTopologyDriven2026}. The resulting mass domain walls can host localized corner modes~\cite{liCreationManipulationHigherorder2024,7q2z-317r,g1cd-9sb5,cpzv-qc5v}.
Recent studies have shown that $d_{x^2-y^2}$-wave altermagnetism can induce higher-order topology~\cite{liMajoranaCornerModes2023a,liCreationManipulationHigherorder2024}. However, the role of $d_{xy}$-wave altermagnetism remains largely unexplored.

In this work, we consider a topological insulator with
proximity-induced $d_{xy}$-wave altermagnetic order. We show that higher-order topological insulator(HOTI) phase emerges for diamond-shaped boundaries, whereas they are absent for square-shaped boundaries. To capture the effects of the boundary geometry, we derive effective edge Hamiltonians for square and diamond boundaries in a common basis. The altermagnetic term has a vanishing projection onto the axial edge, but its projection onto the diagonal edges generates effective masses with alternating signs, giving rise to localized corner modes. Furthermore, zero-energy states localized at $\pi$-flux cores under periodic boundary conditions provide complementary evidence for the higher-order topology. Finally, the combination of the altermagnetic order and the Zeeman field drives a transition between higher-order topological and trivial phases, accompanied by the closing and reopening of the diagonal-edge gap while the bulk gap remains open. These findings establish proximity-induced $d_{xy}$-wave altermagnetism as a promising route to engineering higher-order topology.

\section{Model, symmetries, and bulk spectrum}
We consider a two-dimensional insulator system in proximity to a $d_{xy}$-wave altermagnet, whose Bloch Hamiltonian is described by
\begin{equation}
\begin{aligned}
H(\bm k)&=M(\bm k)\sigma_x+\lambda\sin k_x\sigma_zs_x+\lambda\sin k_y\sigma_zs_y+g(\bm k)\sigma_zs_z,\\
M(\bm k)&=m_0-t\cos k_x-t\cos k_y,\\
g(\bm k)&=J_d\sin k_x\sin k_y.
\end{aligned}
\end{equation}
Here, $s_i$ and $\sigma_i$ are Pauli matrices acting on spin and orbital degrees of freedom, respectively. The nearest-neighbor hopping amplitude is denoted by $t$, while $J_d$ characterizes the effective strength. At $J_d=0$, the isotropic model is a topological insulator for $0<|m_0|<2t$. 

The $d_{xy}$-wave altermagnetic term breaks the conventional time-reversal symmetry. Under time reversal, $\mathcal{T}=is_yK$, the momentum and spin operators transform as $\mathbf{k}\rightarrow-\mathbf{k}$ and $\mathcal{T}s_i\mathcal{T}^{-1}=-s_i$, respectively. The altermagnetic term breaks time-reversal symmetry, as expressed by $\mathcal{T}H(k)\mathcal{T}^{-1}\ne H(-k)$. Nevertheless, the Parity-time $(\mathcal{PT})$ symmetry remains preserved, which protects the twofold degeneracy of the bulk bands. The bulk spectrum is
\begin{equation}
E_\pm(\bm k)=\pm\sqrt{M(k)^2+\lambda^2\sin^2k_x+\lambda^2\sin^2k_y+g(k)^2}.\label{eq:bulk}
\end{equation}

\begin{figure}
\includegraphics[width=0.5\textwidth]{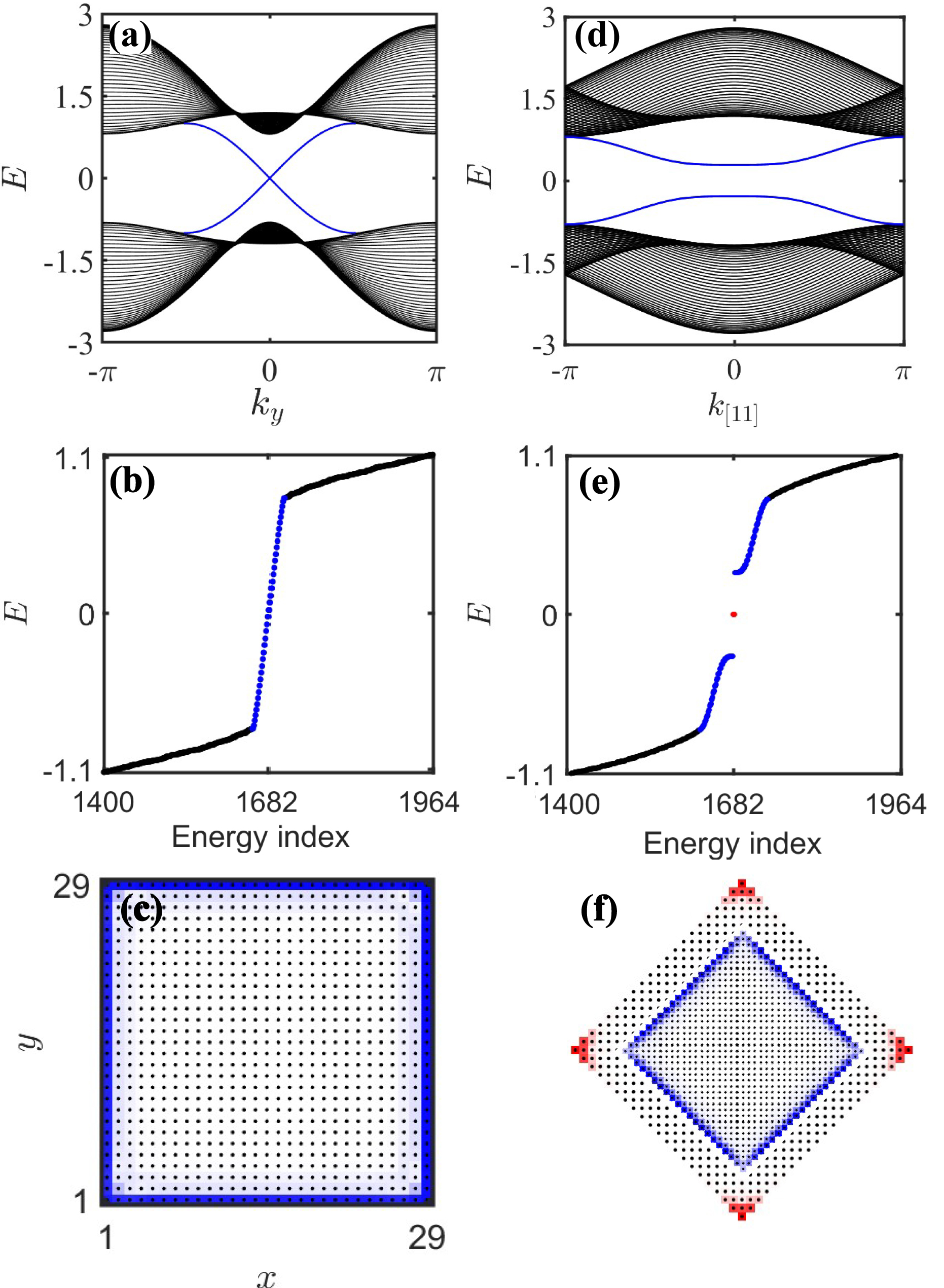}
\caption{Energy spectra and probability distributions under different boundary geometries. 
(a) Energy spectrum of a ribbon geometry as a function of $k_y$ for the system under OBC$_x$-PBC$_y$. 
(b) Energy spectrum of a square lattice under OBC, with the edge states inside the bulk energy gap highlighted in blue.
(c) Average probability density of the edge states under OBC.
(d) Energy spectrum for a ribbon under PBC along the $[11]$ direction and OBC along the $[1\bar{1}]$ direction.
(e) Energy spectrum of a diamond lattice under OBC, with corner modes (red), edge states (blue), and bulk states (black). 
(f) Spatial probability distributions of second-order corner modes, with the inset showing the averaged probability distributions of first-order edge states.
The parameters are $t=1$, $m_0=0.8$, $\lambda=1$, and $J_d=0.6$.}
\label{fig:boundary}
\end{figure}

We compare the energy spectrum and probability distributions of square and diamond lattices.
As shown in Fig.~\ref{fig:boundary}(a), the square lattice with open boundary conditions along the $x$ direction and periodic boundary conditions along the $y$ direction (OBC$_x$-PBC$_y$) exhibits gapless edge states within the bulk energy gap. 
Figure \ref{fig:boundary}(b) shows the energy spectrum of the square lattice under OBC, where the bulk state energies are shown in black and the topological edge state energies are highlighted in blue. The topological edge states are located within the bulk energy gap, and their probability distributions are shown in Fig.~\ref{fig:boundary}(c). These states are mainly distributed along the four edges of the lattice, indicating that the system retains first-order edge states but does not host higher-order corner modes in the square lattice.

When the system is terminated along the diagonal direction, the energy spectrum changes significantly. The altermagnetic term opens an edge-state gap, resulting in four in-gap corner modes. Figure \ref{fig:boundary}(d) shows the ribbon spectrum corresponding to the diamond lattice, with PBC along the $[11]$ direction and OBC along the $[1\bar{1}]$ direction. In contrast to the gapless edge states in Fig.~\ref{fig:boundary}(a), the diagonal edge states become gapped in the presence of altermagnetic order. Figure \ref{fig:boundary}(e) shows the energy spectrum of the diamond lattice under OBC. The spectrum can be classified into bulk states (black), first-order edge states (blue), and second-order corner modes (red). Four corner-mode eigenvalues appear within the edge gap. The corresponding probability distributions are presented in Fig.~\ref{fig:boundary}(f). The higher-order corner modes highlighted in red are strongly localized at the four corners of the diamond lattice, while the inset shows that the first-order edge states highlighted in blue are mainly distributed along the diamond boundaries.

\section{EDGE THEORY}
To elucidate the origin of the higher-order corner modes, we employ edge theory to analyze both square and diamond boundary geometries, and the detailed derivations are given in Appendices~\ref{app:square} and~\ref{app:diamond}. The low-energy Hamiltonian is obtained by expanding the Bloch Hamiltonian to second order in momentum around $(k_x,k_y)=(0,0)$: 

\begin{figure}
\includegraphics[width=0.43\textwidth]{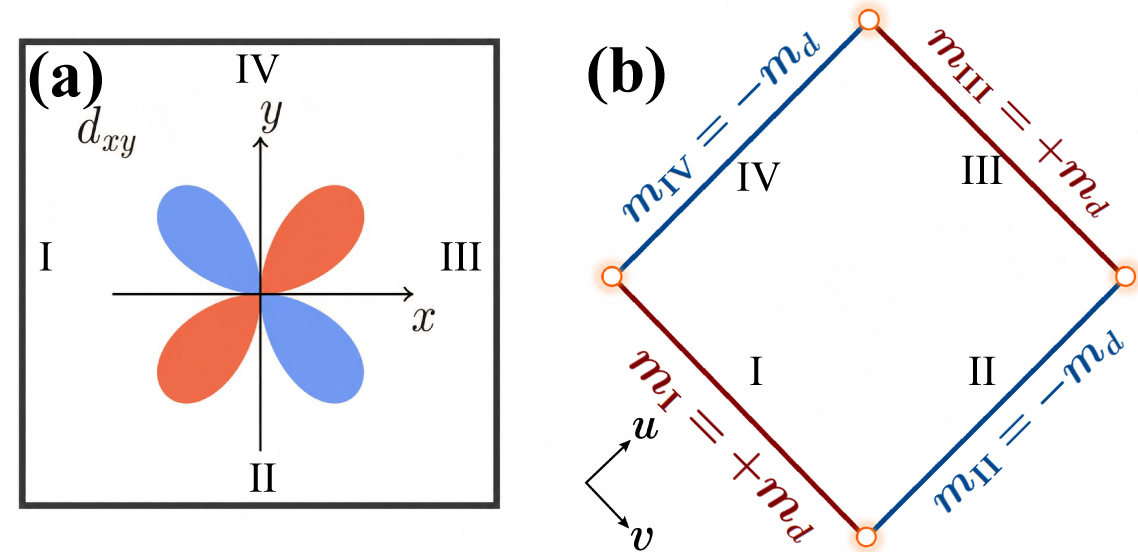}
\caption{Boundary geometry and mass pattern. (a) For the square lattice, the projected effective masses vanish, $m_{\mathrm{I}}=m_{\mathrm{II}}=m_{\mathrm{III}}=m_{\mathrm{IV}}=0$. The inset illustrates the momentum-space structure of the $d_{xy}$ angular form factor. (b) For the diamond lattice, the adjacent edges have opposite mass signs, $m_{\mathrm{I}}=m_{\mathrm{III}}=+m_d$ and $m_{\mathrm{II}}=m_{\mathrm{IV}}=-m_d$, forming mass domain walls at the four corners.}
\label{fig:mass}
\end{figure}

\begin{align}
H(\bm{k})\simeq{}&\left(
m_0-2t+\frac{t}{2}k_x^2+\frac{t}{2}k_y^2\right)\sigma_x\nonumber\\
&+\lambda k_x\,\sigma_zs_x+\lambda k_y\,\sigma_zs_y+J_dk_xk_y\sigma_zs_z.
\end{align}

For the square boundary geometry, we denote the four edges of the square geometry by I--IV, as depicted in Fig.~\ref{fig:mass}(a). For edge I, we substitute $k_x$ with $-i\partial_x$ and neglect the second-order term $k_y^2$, and the resulting Hamiltonian can be divided into two parts, $H=H_0+H_p$, where

\begin{equation}
\begin{aligned}
H_0={}&\left(m_0-2t-\frac{t}{2}\partial_x^2\right)
\sigma_x-i\lambda\partial_x\sigma_zs_x,\\
H_p={}&\lambda k_y\sigma_zs_y-iJ_dk_y\partial_x
\sigma_zs_z.
\end{aligned}
\label{eq:edge_hamiltonian}
\end{equation}

By solving the eigenvalue equation \(H_0|\psi_\alpha\rangle=E_\alpha|\psi_\alpha\rangle\) subject to the boundary conditions \(\psi_\alpha(0)=\psi_\alpha(\infty)=0\), we obtain Jackiw-Rebbi solutions of the form:

\begin{equation}
\psi_{\alpha}(x)
=\mathcal{N}_{x}\sin(\kappa_{1}x)
e^{\kappa_{2}x}e^{ik_y y}\xi_{\alpha},
\end{equation}
where $\kappa_{1}=\frac{\sqrt{-\lambda^2-2Mt}}{t}$ and $\kappa_{2}=-\frac{\lambda}{t}$. The normalization factor is $\mathcal{N}_x=\sqrt{\frac{-4\kappa_2(\kappa_1^2+\kappa_2^2)}{\kappa_1^2}}$. The eigenvector $\xi_\alpha$ satisfies $\sigma_ys_x\xi_\alpha=\xi_\alpha$ and can be explicitly expressed as

\begin{equation}
\begin{aligned}
\xi_1 &= |\sigma_y=+1\rangle \otimes |s _x=+1\rangle,\\
\xi_2 &= |\sigma_y=-1\rangle \otimes |s _x=-1\rangle.
\end{aligned}
\end{equation}

The effective Hamiltonian for edge I is obtained by projecting $H_p$ onto the edge-state basis, $H_{\mathrm{edge}}^{\mathrm{I},\alpha\beta}(k_y)=\int_0^\infty dx\,\psi_\alpha^*(x)H_p(-i\partial_x,k_y)\psi_\beta(x)$. The effective Hamiltonians for the remaining three edges can be derived analogously and are expressed as

\begin{equation}
\begin{aligned}
H_{\mathrm{I}}
=-\lambda k_y\tau_y,\\
H_{\mathrm{II}}
=+\lambda k_x\tau_y,\\
H_{\mathrm{III}}
=+\lambda k_y\tau_y,\\
H_{\mathrm{IV}}
=-\lambda k_x\tau_y.
\end{aligned}
\end{equation}

We define an edge coordinate $l$ in the anticlockwise direction. The effective edge Hamiltonian can be expressed as
\begin{equation}
H_{\mathrm{edge}}
=
-i\lambda\tau_y\partial_{l}.
\label{eq:edge_hamiltonian}
\end{equation}

For the diamond boundary geometry, we denote the four edges of the diamond geometry by I--IV, as depicted in Fig.~\ref{fig:mass}(b). The rotated momenta and spin matrices are defined as $k_{u,v}=(k_x\pm k_y)/2$ and $s_{\pm}=(s_x\pm s_y)/\sqrt{2}$, respectively. We first consider edge IV, $k_u$ and $k_v$ denote the momenta parallel and perpendicular to the boundary, respectively. Expanding the Hamiltonian to first order in $k_u$, we write $H^d=H_{0}^d+H_{p}^d$, where

\begin{equation}
\begin{aligned}
H_{0}^d={}&(m_0-2t\cos k_v)\sigma_x
+\sqrt{2}\lambda\sin k_v\sigma_zs_-,
\\
H_{p}^d={}&\sqrt{2}\lambda k_u\cos k_v\sigma_zs_+
-J_d\sin^2k_v\sigma_zs_z.
\end{aligned}
\end{equation}

To retain the lattice contributions along the normal direction, we solve the eigenvalue equation $H_0^d|\psi_d\rangle=E_d|\psi_d\rangle$. The corresponding localized wave functions take the form $\psi_d(n)=\mathcal{N}(z_1^n-z_2^n)\chi$, where $\mathcal{N}$ is the normalization factor and $z_{1,2}$ are the two decaying roots ($|z_{1,2}|<1$) satisfying:

\begin{equation}
\left(t+\frac{\lambda}{\sqrt{2}}\right)z^2
-m_0z+\left(t-\frac{\lambda}{\sqrt{2}}\right)=0.
\end{equation}

Considering the boundary condition, the eigenvectors $\chi$ satisfy $\sigma_ys_-\chi=\chi$ and can be chosen as
\begin{equation}
\begin{aligned}
\chi_1&=
|\sigma_y=+1\rangle\otimes|s_-=+1\rangle,\\
\chi_2&=
|\sigma_y=-1\rangle\otimes|s_-=-1\rangle.
\end{aligned}
\end{equation}

Consequently, the effective Hamiltonians for the remaining three edges can be derived analogously and are expressed as
\begin{equation}
\begin{aligned}
H_{\mathrm{I}}^d
&=v_dk_{v}\tau_y
+m_d\tau_x,\\
H_{\mathrm{II}}^d
&=v_dk_u\tau_y
-m_d\tau_x,\\
H_{\mathrm{III}}^d
&=-v_dk_v\tau_y
+m_d\tau_x,\\
H_{\mathrm{IV}}^d
&=-v_dk_u\tau_y
-m_d\tau_x.
\end{aligned}
\end{equation}
where $v_d=\frac{\sqrt{2}\lambda m_0}{2t}$ and $m_d=\frac{4t^2-m_0^2}{4t(t+\lambda/\sqrt{2})}J_d$. We define an edge coordinate $l$ in the anticlockwise direction. The effective edge Hamiltonian can be expressed as
\begin{equation}
H_{\mathrm{edge}}^d
=
-iv_d\tau_y\partial_{l}+M_d(l)\tau_x,
\end{equation}
with $M_d(l)=+m_d$, $-m_d$, $+m_d$ and $-m_d$ for edges I, II, III, and IV, respectively. The above edge theory reveals a distinct dependence of higher-order topology on the boundary geometry. For the square geometry, the $d_{xy}$-wave altermagnetic term has a vanishing projection onto the edge-state subspace, resulting in zero effective masses and the absence of higher-order corner modes.  In contrast, for the diamond geometry, the altermagnetic term generates finite effective masses with alternating signs, $m_{\mathrm{I}}=m_{\mathrm{III}}=m_d$ and $m_{\mathrm{II}}=m_{\mathrm{IV}}=-m_d$. The sign change in the effective masses gives rise to Jackiw--Rebbi domain walls at the four corners, which bind localized corner modes. These results demonstrate that the emergence of higher-order topological corner modes depends on the boundary geometry.
\begin{figure}[t]
\includegraphics[width=\columnwidth]{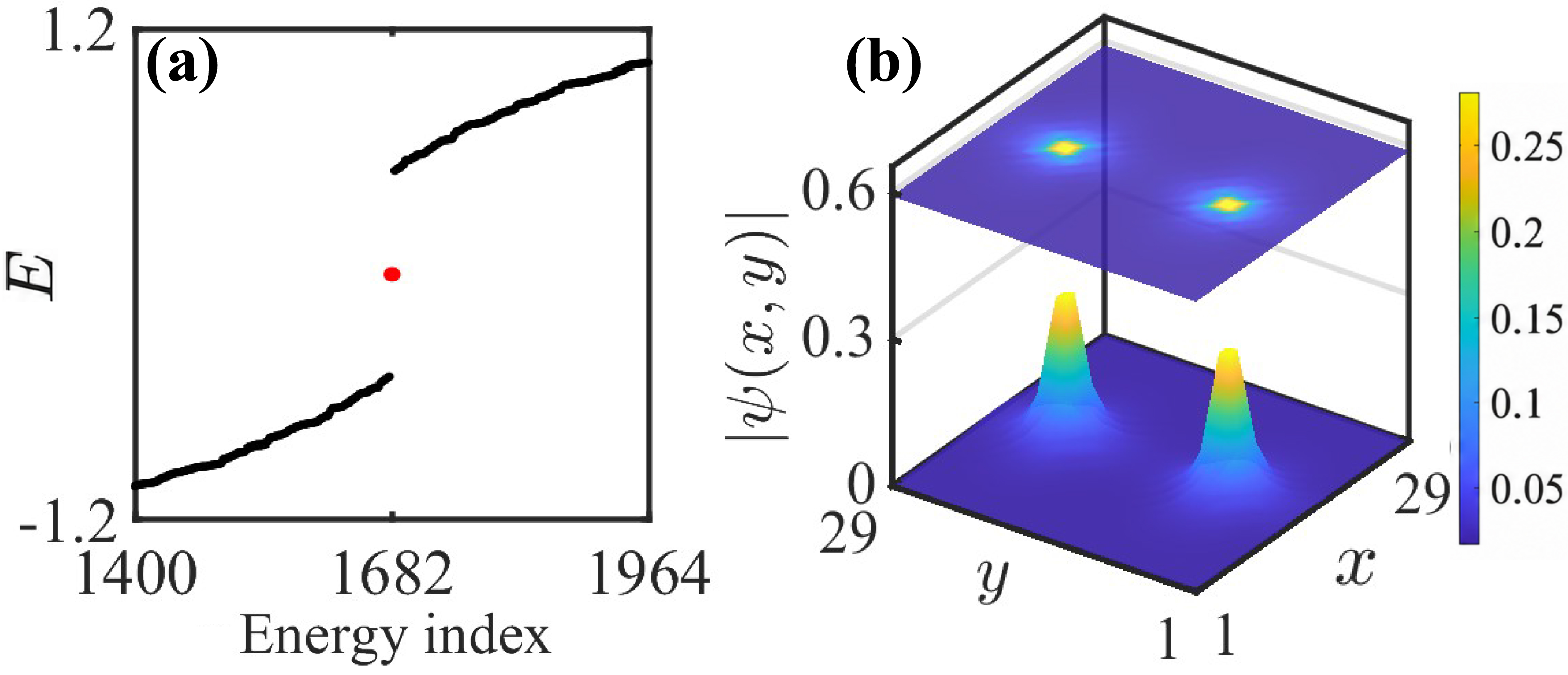}
\caption{(a) Energy spectrum of the finite lattice in the presence of two $\pi$-flux defects under PBC, with the $\pi$-flux induced zero-energy states highlighted in red. The other eigenstates are shown in black. (b) Probability density of the corresponding zero-energy state, and its projection onto the $x-y$ plane. The parameters are set as  $t=1$, $m_0=0.8$, $\lambda=1$, and $J_d=0.6$.}
\label{fig:defects}
\end{figure}

To further investigate the higher-order topological properties of the system, we insert two $\pi$ fluxes by choosing a branch cut connecting their cores and multiplying the hopping amplitudes crossing the cut by $-1$, which is equivalent to introducing a Peierls phase shift of $\pi$. This construction includes diagonal bonds as well as nearest-neighbor bonds. Although the choice of the branch cut is gauge dependent, the enclosed flux is gauge invariant. PBC eliminates physical edge and corner boundaries, allowing the localized defect response to be examined separately. This configuration enables us to distinguish flux-induced bound states from conventional boundary states.

We numerically diagonalize the Hamiltonian under PBC, and the resulting energy spectrum is shown in Fig.~\ref{fig:defects}(a). Four zero-energy states appear inside the bulk gap and are clearly separated from the bulk bands. To characterize the spatial localization of these states, we calculate the probability density distributions of the corresponding eigenstates. As shown in Fig.~\ref{fig:defects}(b), the probability density is strongly localized around the two $\pi$-flux cores and rapidly decays away from the defect positions. This spatial localization indicates that the in-gap states are bound to the flux defects rather than extended throughout the bulk.

The zero-energy states localized at the $\pi$-flux cores reveal a nontrivial defect response under PBC. While the edge theory explains the emergence of corner modes through alternating Dirac masses, the $\pi$-flux calculation provides complementary evidence for the nontrivial topology without introducing physical boundaries.

\section{Zeeman-field driven topological transition in $d_{xy}$ altermagnet}
\begin{figure*}
\includegraphics[width=0.8\textwidth]{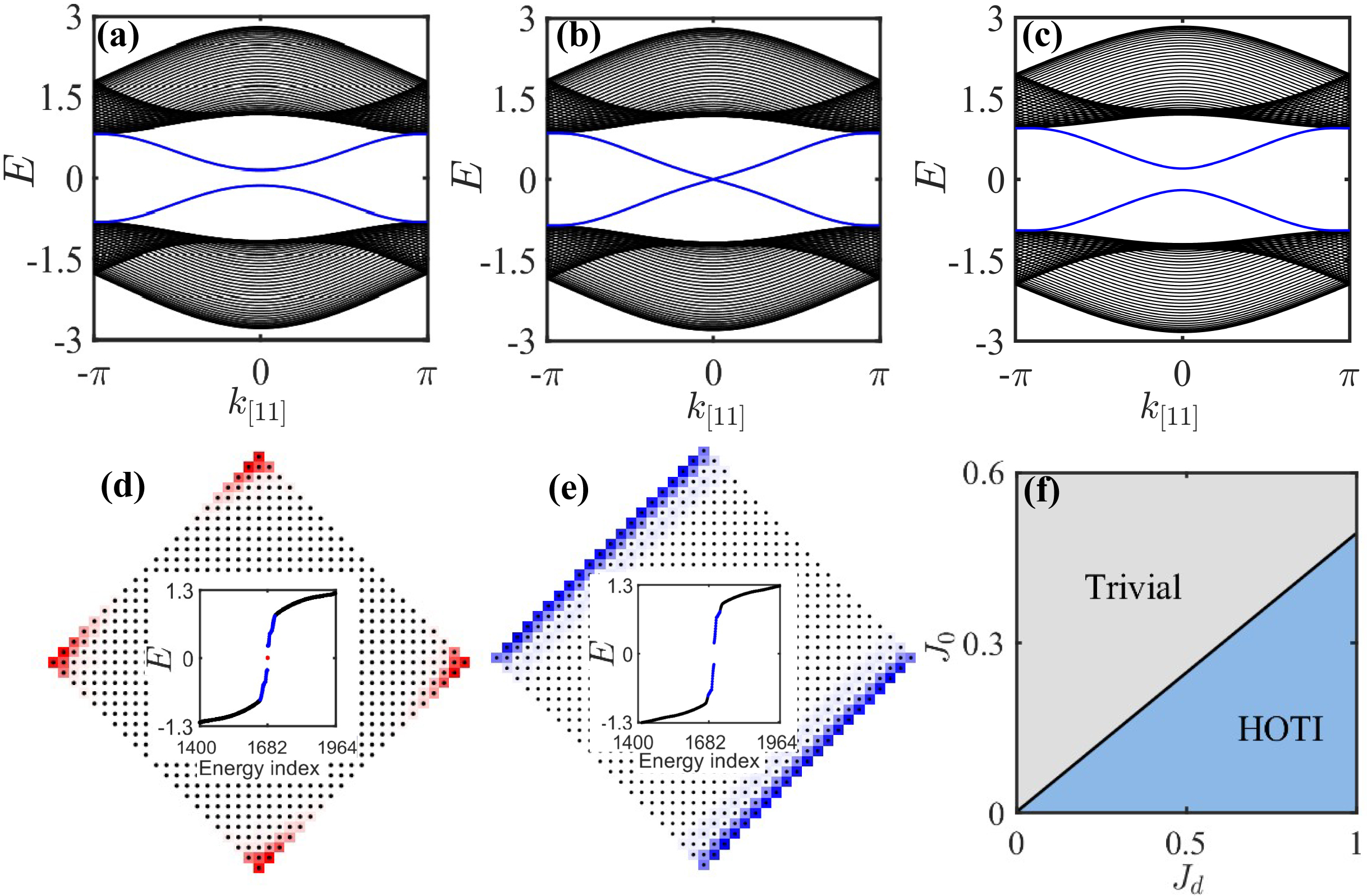}
\caption{
(a)-(c) Energy spectra for a ribbon geometry corresponding to the diagonal edges of the system, plotted as a function of $k_{[11]}$ for $J_0=0.15$, $J_c$, and $0.5$, respectively.
(d) Probability distributions of second-order corner modes for $J_0=0.15$. The inset shows energy spectrum of a diamond lattice under OBC, with corner modes (red), edge states (blue), and bulk states (black).
(e) Probability distributions of edge states for $J_0=0.5$. The inset shows the energy spectrum of the diamond lattice under OBC, with edge states (blue) and bulk states (black).
(f) Phase diagram in the $J_d-J_0$ plane for the diamond lattice. Other parameters are $t=1$, $m_0=0.8$, $J_d=0.6$, and $\lambda=1$.}
\label{fig:J0}
\end{figure*}

After establishing the higher-order corner modes induced by $J_d$, we further investigate the influence of the Zeeman field $J_0$ on the edge gap and corner modes, and whether it can drive a transition between the HOTI and trivial phases. The Bloch Hamiltonian is written as follows:
\begin{equation}
\begin{aligned}
H(\bm{k})=\left(m_0-t\cos k_x-t\cos k_y\right)\sigma_x+\lambda\sin k_x\,\sigma_z s_x\\
+\lambda\sin k_y\,\sigma_z s_y+\left(J_0+J_d\sin k_x\sin k_y
\right)\sigma_z s_z .
\end{aligned}
\end{equation}
As derived in Appendix~\ref{app:diamond}, the edge spectrum along the $u$ direction is given by
\begin{equation}
E_{\pm}(k_u)
=\pm\sqrt{v_d^2k_u^2+(J_0-m_d)^2},
\end{equation}
where $m_d=\frac{4t^2-m_0^2}{4t(t+\lambda/\sqrt2)}J_d$. Within this leading Dirac description, the minimum edge gap is determined by
\begin{equation}
\Delta_{\mathrm{edge}}
=2||J_0|-|m_d||.
\end{equation}

The effective masses associated with the two inequivalent edge orientations of the diamond geometry become $J_0+m_d$ and $J_0-m_d$, respectively. For $J_0<m_d$, the masses of adjacent edges have opposite signs, giving rise to Jackiw--Rebbi mass domain walls that support localized corner modes. In contrast, for $J_0>m_d$, the adjacent effective masses have the same sign, and the corner modes disappear. For $t=1, \lambda=1$, the critical Zeeman field is determined by $J_c=m_d$ and can be simplified as
\begin{equation}
J_c=\frac{2-\sqrt{2}}{4}(4-m_0^2)J_d.
\end{equation}

Figures \ref{fig:J0}(a)-–(c) show the ribbon spectra of the diamond lattice for $J_0=0.15$, $J_c$, and $0.5$, respectively. The altermagnetic strength fixed at $J_d=0.6$. For $J_0=0.15$, the energy spectrum exhibits a finite gap around zero energy, as shown in Fig.~\ref{fig:J0}(a).
As the Zeeman field $J_0$ increases, the edge gap gradually decreases and eventually closes at the critical value $J_0=J_c$, as shown in Fig.~\ref{fig:J0}(b). For $J_0=0.5$, the edge gap reopens after passing through the critical point, as shown in Fig.~\ref{fig:J0}(c). Throughout this evolution, the bulk gap remains open, indicating a topological phase transition driven by the closing and reopening of the edge gap.

For $J_0 < m_d$, the adjacent edges have opposite mass signs, preserving the mass domain walls and supporting corner modes. For $J_0>J_c$, the adjacent edges acquire the same sign, the mass domain walls at the corners disappear, and the higher-order corner modes are no longer supported. Therefore, although the bulk gap remains open, the closing and reopening of the edge gap signal a transition from the higher-order topological phase to the trivial phase.

As shown in Fig.~\ref{fig:J0}(d), the corner modes highlighted in red are strongly localized at the four corners of the diamond lattice, and the inset shows the corresponding energy spectrum under OBC. In contrast, the edge states are mainly distributed along the $[11]$ boundaries of the diamond lattice for $J_0=0.5$, as shown in Fig.~\ref{fig:J0}(e). The inset shows the energy spectrum under OBC, with no corner-state eigenvalues remaining within the edge gap, while the remaining spectrum consists of bulk and edge states (blue). Unlike the corner modes in the HOTI phase, these edge states are not topologically protected and are sensitive to perturbations. Figure \ref{fig:J0}(f) presents the phase diagram in the $J_d-J_0$ plane. The phase boundary is determined by the sign change of the effective mass terms in the edge theory and is given by $J_0=\frac{2-\sqrt{2}}{4}(4-m_0^2)J_d$, which separates the parameter space into the HOTI and the trivial phase.   

\section{conclusions and Discussion }
In this work, we have investigated higher-order topology induced by proximity to a $d_{xy}$-wave altermagnet in a two-dimensional topological insulator. In the absence of a Zeeman field, the altermagnetic term has a vanishing projection onto the axis-aligned edge-state subspace, leaving the edge states gapless. In contrast, diagonal edges acquire effective masses with alternating signs. The resulting mass domain walls support four corner-localized modes. In addition, the zero-energy states localized at $\pi$-flux cores under PBC provide complementary evidence for the higher-order topology without introducing physical boundaries. the combination of the altermagnetic term and the Zeeman field enables a transition between HOTI and trivial phases through the closing and reopening of the edge gap, while the bulk gap remains open. Our results reveal the crucial role of boundary geometry in determining higher-order topology, and offer a promising route to engineering topological corner modes through altermagnetism.

From an experimental perspective, a promising route is to construct a heterostructure combining a two-dimensional topological insulator with a candidate $d_{xy}$-wave altermagnet. The effective altermagnetic strength can be tuned by varying the interlayer distance. Altermagnetic proximity effects have been theoretically predicted in V$_2$Se$_2$O-based heterostructures, where the induced spin splitting can be tuned by interlayer spacing and magnetic configuration~\cite{zhuAltermagneticProximityEffect2026}. The altermagnetic order in our model can be realized in heterostructures by interfacing a two-dimensional topological insulator with a layered altermagnet. The predicted corner modes could be probed using scanning tunneling microscopy and spectroscopy~\cite{PhysRevB.108.085422,huIdentifyingTopologicalCorner2023}, which could reveal in-gap resonances localized at the corners and distinct spectral features along edges with different orientations. 

\begin{acknowledgments}
This study received support from the National Natural Science Foundation of China (Grant No. 12574266), National Key Research and Development Program of China (Grant No. 2022YFA1405200) and the Open Research Project of the National Key Laboratory of Infrared Science and Technology (SITP-SKLIP-YB-2025-09).
\end{acknowledgments}
\section*{Data availability}
The data that support the findings of this article are not publicly available. The data are available from the authors
upon reasonable request.

\appendix
\section{Edge theory for square boundaries}
\label{app:square}
Expand the Bloch Hamiltonian in momentum around $(k_x,k_y)=(0,0)$ and set $M=m_0-2t$. We consider the general form of the introduced Zeeman field $J_0$ and substitute $k_x$ with $-i\partial_x$. For the edge I, the low-energy Hamiltonian is described by:
\begin{align}
H&=H_{0}+H_{p},\nonumber\\
H_{0}&=(M-\tfrac{t}{2}\partial_x^2)\sigma_x
 -i\lambda\partial_x\sigma_z s_x,\label{eq:squareH0}\\
H_{p}&=\lambda k_y\sigma_zs_y
 +(J_0-iJ_dk_y\partial_x)\sigma_z s_z.
\label{eq:squareH1}
\end{align}
Here, $H_{0}$ is solved at zero longitudinal momentum, and $H_{p}$ is projected to first order. The boundary conditions impose $\psi(0)=\psi(\infty)=0$.

By solving the eigenvalue equation $H_0|\psi\rangle=E|\psi\rangle$ and multiply by $\sigma_x$. Since $\sigma_x\sigma_z=-i\sigma_y$,
\begin{equation}
\left[M-\tfrac{t}{2}\partial_x^2-\lambda\partial_x(\sigma_ys_x)\right]\psi=0.
\end{equation}
Using a decaying ansatz, 
\begin{align}
\psi=e^{\kappa x}\xi_\alpha,\quad \sigma_ys_x\xi_\alpha=\eta\xi_\alpha,\quad \eta=\pm1,
\end{align}
the zero-energy equation gives
\begin{align}
&(M-\tfrac{t}{2}\partial_x^2-\lambda\partial_x)\psi=0,\nonumber\\
&\kappa_{x,\pm}=\frac{-\eta \lambda\pm\sqrt{\lambda^2+2tM}}{t}.
\label{eq:square-roots}
\end{align}
For $\eta=1$, the two roots have negative real parts, allowing a normalizable edge-state solution. We obtain  the Jackiw-Rebbi solutions
\begin{align}
\psi_{\alpha}(x)
=\mathcal{N}_{x}\sin(\kappa_{1}x)
e^{\kappa_{2}x}e^{ik_y y}\xi_{\alpha}.
\label{eq:normalized-envelope}
\end{align}
There, $\kappa_{1}=\frac{\sqrt{-\lambda^2-2Mt}}{t}$ and $\kappa_{2}=-\frac{\lambda}{t}$. The normalization factor is $\mathcal{N}_x=\sqrt{\frac{-4\kappa_2(\kappa_1^2+\kappa_2^2)}{\kappa_1^2}}$.

For an inward normal at angle $\theta$, choose $\eta=\pm1$ and write the normalized orbital spinors as
\begin{align}
|\sigma_y=\eta\rangle=\frac{1}{\sqrt2}\begin{pmatrix}1\\ i\eta\end{pmatrix},\quad
|s_n=\eta\rangle=\frac{1}{\sqrt2}
\begin{pmatrix}1\\\eta e^{i\theta}\end{pmatrix}.
\end{align}
A convenient
explicit choice is
\begin{align}
\xi_1 &= |\sigma_y = +1\rangle \otimes |s_n = +1\rangle, \nonumber\\
\xi_2 &= |\sigma_y = -1\rangle \otimes |s_n = -1\rangle.
\end{align}
Having fixed the edge basis, the effective one-dimensional
Hamiltonian follows by projecting the tangential and altermagnetic terms onto this subspace. Let
\begin{align}
P^\dagger\left(\sigma_zs_y\right)
P=-\tau_y,\nonumber\\
P^\dagger\left(\sigma_zs_z\right)
P=\tau_x.
\end{align}

Thus, the effective Hamiltonian for edge I is $H_{\mathrm{I},\alpha\beta}(k_y)
=\int_{0}^{\infty} dx\,
\psi_\alpha^{*}(x)
H_p(-i\partial_x,k_y)
\psi_\beta(x)$. 
Integration by parts gives
\begin{align}
\langle\partial_x\rangle
&=\int_{0}^{\infty}f(x)\partial_ xf(x)dx=\tfrac12[f_x^2]_{0}^{\infty}=0.\nonumber\\
\label{eq:moments}
\end{align}
The projected effective Hamiltonian for edge I is
\begin{align}
H_{edge}^{I}=-\lambda k_y\tau_y+J_0\tau_x.
\end{align}
The effective Hamiltonians for the remaining three edges can be obtained analogously by changing the inward normal direction. With the boundary coordinate $l$ increasing counterclockwise, the corresponding tangential momenta are $-k_y$, $k_x$, $k_y$, and $-k_x$. The effective edge Hamiltonian takes the unified form
\begin{equation}
H_{\mathrm{edge}}
=-i\lambda\tau_y\partial_l+J_0\tau_x,
\end{equation}
For $J_0=0$, the effective masses vanish, and the edge states remain gapless. For finite $J_0$, all four edges acquire the same mass. Consequently, no mass domain walls form at the corners, and the corresponding corner modes are absent.

\section{Edge theory  for diagonal diamond boundaries}
\label{app:diamond}
We first rotate the momentum coordinates and then solve the corresponding one-dimensional boundary problem. Define $k_u=(k_x+k_y)/2$, $k_v=(k_x-k_y)/2$, and $s_\pm=(s_x\pm s_y)/\sqrt2$. The Bloch Hamiltonian is
\begin{align}
H^d={}&(m_0-2t\cos k_u\cos k_v)\sigma_x\nonumber\\
&+\sqrt2 \lambda\sin k_u\cos k_v\,\sigma_zs_+\nonumber\\
&+\sqrt2 \lambda\cos k_u\sin k_v\,\sigma_zs_-\nonumber\\
&+[J_0+J_d(\sin^2k_u-\sin^2k_v)]\sigma_zs_z.
\label{eq:latticeuv}
\end{align}
We consider edge IV, where $k_u$ is the conserved tangential momentum and $k_v$ is the momentum perpendicular to the boundary. The Hamiltonian
is separated into
\begin{equation}
\begin{aligned}
H^d &= H_0^d+H_p^d,\\
H_0^d &= (m_0-2t\cos k_v)\sigma_x
+\sqrt{2}\lambda\sin k_v\,\sigma_z s_-,\\
H_p^d &= \sqrt{2}\lambda k_u\cos k_v\,\sigma_z s_+
+(J_0-J_d\sin^2 k_v)\sigma_z s_z.
\end{aligned}
\end{equation}
The layers have physical separation $1/\sqrt2$. The zero-energy boundary states are obtained from
\begin{align}
H_0^d\psi_d=0.
\end{align}
Using $\cos k_v=(e^{ik_v}+e^{-ik_v})/{2}$, $\sin k_v=(e^{ik_v}-e^{-ik_v})/{2i}$, the Hamiltonian can be written as
\begin{align}
H_0^d=\sum_n C_n^\dagger m_0\sigma_x C_n+\sum_n
(C_n^\dagger T C_{n+1}+H.c.).
\end{align}
The transverse matrix element connecting adjacent layers is
\begin{align}
T=-t\sigma_x-\frac{i\lambda}{\sqrt{2}}\sigma_zs_- .
\end{align}
We take the wave function as $\psi_n=f_n\chi$,
where the spinor satisfies $\sigma_ys_-\chi=\eta\chi,
\eta=\pm1$.
The zero-energy equation gives the recurrence relation
\begin{align}
\left(t+\eta\frac{\lambda}{\sqrt2}\right)f_{n+1}-m_0f_n+\left(t-\eta\frac{\lambda}{\sqrt2}\right)f_{n-1}=0 .
\label{A4}
\end{align}
Introducing the ansatz $f_n=z^n$, we obtain
\begin{equation}
\begin{aligned}
&\left(t+\eta\frac{\lambda}{\sqrt2}\right)z^2-m_0z+\left(t-\eta\frac{\lambda}{\sqrt2}\right)=0,\\
&z_{1,2}=\frac{m_0\pm\sqrt{m_0^2-4(t^2-\lambda^2/2)}}{2(t+\eta \lambda/\sqrt2)}.
\end{aligned}
\end{equation}
For the edge-localized solution, $|z|<1$ requires $\eta=1$.

Write $c_\ell=\sum_{n\ge1}f_nf_{n+\ell}$. Equation~\eqref{A4} can be rewritten as $f_{n+1}=\alpha f_n-\beta f_{n-1}$, where $\alpha=\frac{m_0}{t+\lambda/\sqrt2}, \beta=\frac{t-\lambda/\sqrt2}{t+\lambda/\sqrt2}$. Multiplication by $f_n$ and summation,
\begin{align}
c_1&=\alpha-\beta c_1=\frac{m_0}{2t},\nonumber\\
c_2&=\alpha c_1-\beta
=\frac{m_0^2}{2t(t+\lambda/\sqrt2)}-\frac{t-\lambda/\sqrt2}{t+\lambda/\sqrt2}.
\label{eq:corr}
\end{align}
The two zero-energy edge basis states are chosen as
\begin{equation}
\begin{aligned}
\chi_+
&=
|\sigma_y=+1\rangle
\otimes
|s_-=+1\rangle ,
\\
\chi_-
&=
|\sigma_y=-1\rangle
\otimes
|s_-=-1\rangle .
\end{aligned}
\end{equation}
Using these basis states, the projected Pauli matrices satisfy
\begin{equation}
\begin{aligned}
P^\dagger \sigma_z s_{+}P=-\tau_y,\qquad
P^\dagger\sigma_zs_zP=\tau_x .
\end{aligned}
\end{equation}

The term $\sin^2 k_v$ in the bulk Hamiltonian can be represented in the
perpendicular lattice direction as
\begin{equation}
\sin^2 k_v
=\frac{1}{2}
-\frac{1}{4}\left(
e^{2ik_v}+e^{-2ik_v}
\right).
\end{equation}
Projecting this expression onto the normalized edge-state wave function yields
\begin{equation}
\begin{aligned}
&\sum_{n\geq1}f_n
\left[\frac12f_n
-\frac14(f_{n+2}+f_{n-2})\right]=\frac12(1-c_2).
\end{aligned}
\end{equation}
The altermagnetic contribution to the effective mass is given by $m_d=\frac{(1-c_2)}{2}J_d=\frac{4t^2-m_0^2}{4t(t+\lambda/\sqrt{2})}J_d$.
The projected effective Hamiltonian for edge IV is
\begin{equation}
H_{\mathrm{IV}}^d=-v_dk_{u}\tau_y+(J_0-m_d)\tau_x.  
\end{equation}
 
With the boundary coordinate $l$ increasing counterclockwise, the corresponding tangential momenta are $k_v$, $k_u$, $-k_v$, and $-k_u$. The four edge Hamiltonians can be written in the unified form
\begin{equation}
H_{\mathrm{edge}}^{d}
=-iv_d\tau_y\partial_l+M_d(l)\tau_x,
\end{equation}
where $M_d(l)=J_0+m_d$, $J_0-m_d$, $J_0+m_d$ and $J_0-m_d$ for edges I, II, III, and IV, respectively.

In the absence of the Zeeman field, the effective masses alternate in sign between adjacent edges, giving rise to mass domain walls at the four corners.
For $J_0<m_d$, the mass domain walls persist and support localized corner modes. At $J_0=m_d$, the masses on edges II and IV vanish. For $J_0>m_d$, all four effective masses have the same sign, and the corner states disappear.

\bibliographystyle{apsrev4-2}
% Write the style-only control citation without asking natbib for a numbered item.
\makeatletter
\immediate\write\@auxout{\string\citation{prbTitleControl}}
\makeatother
\bibliography{references}
\end{document}